\documentclass[]{cernrep}
\usepackage{graphicx}
\usepackage{epsfig}
\usepackage{rotating}
\usepackage{dcolumn}
\usepackage{bm}

\newcommand{\lsim}
{{\;\raise0.3ex\hbox{$<$\kern-0.75em\raise-1.1ex\hbox{$\sim$}}\;}}
\newcommand{\gsim}
{{\;\raise0.3ex\hbox{$>$\kern-0.75em\raise-1.1ex\hbox{$\sim$}}\;}}

\title{The derivative of the topological susceptibility at zero momentum and an
estimate of $\eta'$ mass in the chiral limit}
\author{J.~Pasupathy$^a$, J.~P.~Singh$^b$, R.~K.~Singh$^a$ and  
A.~Upadhyay$^b$\\[0.2cm]
$^a$ Center for High Energy Physics, Indian Institute of Science, 
Bangalore-560 012, India\\
$^b$ Physics Department, Faculty of Science, M.S. University of Baroda, 
Vadodara-390 002, India}
\begin{document}
\maketitle
\begin{abstract}
The anomaly-anomaly correlator is studied using QCD sum rules. Using the matrix
elements of anomaly between vacuum and pseudoscalars $\pi, \ \eta$ and $\eta'$,
the derivative of correlator $\chi'(0)$ is evaluated and found to be $\approx 
1.82 \times 10^{-3}$ GeV$^2$. Assuming that $\chi'(0)$ has no significant
dependence on quark masses, the mass of $\eta'$ in the chiral limit is found to
be $\approx$723 MeV. The same calculation also yields for the singlet
pseudoscalar decay constant in the chiral limit a value of $\approx 178$ MeV.
\end{abstract}
%
%
In QCD, it is known that the vacuum gauge field configurations are significant,
in particular the instanton solutions corresponding to self-dual fields
$(G^a_{\mu\nu} = \pm (1/2) \tilde{G}^a_{\mu\nu})$ play a role in solving the so
called $U(1)$ problem, i.e., the ninth pseudoscalar $\eta'$ remains massive even
in the chiral limit when all quark masses are zero. 
The axial vector current in QCD has an anomaly
\begin{equation}
\partial^\mu\bar{q}\gamma_\mu\gamma_5 q = 2 \ i \ m_q \ \bar{q}\gamma_5 q -
\frac{\alpha_s}{4\pi}  \ G^a_{\mu\nu}\tilde{G}^{a\mu\nu},\hspace{0.5cm}
\mbox{where,} \hspace{0.2cm}\tilde{G}^{a\mu\nu} = \frac{1}{2}
\epsilon^{\mu\nu\rho\sigma}G^a_{\rho\sigma}.
\end{equation}
The topological susceptibility $\chi(q^2)$ defined by
\begin{equation}
\chi(q^2) = i \intop d^4x \ e^{iq.x} \langle0|T\left\{ Q(x), \ Q(0)
\right\}|0\rangle, \hspace{0.5cm}\mbox{with,}\hspace{0.2cm}
Q(x) = \frac{\alpha_s}{8\pi} \ G^a_{\mu\nu}\tilde{G}^{a\mu\nu}
\label{chi}
\end{equation}
is of considerable theoretical interest and has been studied using a variety of
theoretical tools like latice guage theory, QCD sum rules, chiral perturbation
theory etc. In particular the derivative of the susceptibility at $q^2=0$
\begin{equation}
\chi'(0) = \frac{d\chi(q^2)}{dq^2}\Big|_{q^2=0}
\end{equation}
enters in the discussion of the proton-spin
problem~\cite{narison,ayk,ago,bli,avs}. In the QCD sumrule approach, one can
determine $\chi'(0)$ as follows. Using despersion relation one can write
\begin{equation}
\frac{\chi'(q^2)}{q^2} - \frac{\chi'(0)}{q^2} = \frac{1}{\pi} \intop ds \ 
\Im(\chi(s)) \left[\frac{1}{s(s-q^2)^2} + \frac{1}{s^2(s-q^2)}\right] + 
\mbox{subtractions}.
\label{dis}
\end{equation}
Defining the Borel transform of a fuction $f(q^2)$ by
\begin{equation}
\hat{B} f(q^2) = \stackrel{Lim}{-q^2,n\to\infty} \ \left[
\frac{(-q^2)^{n+1}}{n!} \ \left(\frac{d}{dq^2}\right)^n \ f(q^2)\right]_{-q^2/n=M^2}
\end{equation}
one gets from Eq.(\ref{dis})
\begin{equation}
\chi'(0) = \frac{1}{\pi} \intop ds \ \frac{\Im(\chi(s)}{s^2}\left(1+\frac{s}{M^2}
\right)\ e^{-s/M^2} - \hat{B}\left[\frac{\chi'(q^2)}{q^2}\right].
\label{chip0}
\end{equation}
According to Eq.(\ref{chi}), $\Im(\chi(s))$ receives contribution from all
states $|n\rangle$ such that $\langle0|Q|n\rangle\neq0$. In particular we
have~\cite{dig}
\begin{equation}
\langle0|Q|\pi^0\rangle = i \ f_\pi \ m_\pi^2 \ \left(\frac{m_d-m_u}{m_d+m_u} 
\right) \frac{1}{2\sqrt{2}}.
\label{vevQ}
\end{equation}
The matrix elements, when $|n\rangle$ is $|\eta\rangle$ or $|\eta'\rangle$, can
be determined as follows. It is known from both theoretical considerations based
on chiral perturbation theory as well as phenomenological analysis that one
needs two mixing angles $\theta_8$ and $\theta_0$ to describe the coupling of the
octet and siglet axial vector currents to $\eta$ and $\eta'$~\cite{hl,feld,re}.
Introduce the definition
\begin{equation}
\langle0|J^a_{\mu5}|P(p)\rangle = i \ f^a_P \ p_\mu; \ a=0,8; \ P=\eta,\eta',
\end{equation}
where $J^{8,0}_{\mu5}$ are the octet and singlet axial currents :
\begin{eqnarray}
J^8_{\mu5} &=& \frac{1}{\sqrt{6}} \left( \bar{u}\gamma_\mu\gamma_5 u +
 \bar{d}\gamma_\mu\gamma_5 d - 2  \bar{s}\gamma_\mu\gamma_5 s \right)\\
J^0_{\mu5} &=& \frac{1}{\sqrt{3}} \left( \bar{u}\gamma_\mu\gamma_5 u +
 \bar{d}\gamma_\mu\gamma_5 d +  \bar{s}\gamma_\mu\gamma_5 s \right).
\end{eqnarray}
The $|P(p)\rangle$ represents either $\eta$ or $\eta'$ with momentum $p_\mu$.
The couplings $f^a_P$ can be equivalently represented by two couplings $f_8, \ 
f_0$ and two mixing angles $\theta_8$ and $\theta_0$ by the matrix identity
\begin{eqnarray}
\left(\begin{tabular}{cc}
$f^8_\eta$ & $f^0_\eta$ \\ $f^8_{\eta'}$ & $f^0_{\eta'}$ \end{tabular} \right)
= \left(\begin{tabular}{cc}
$f_8 \ \cos\theta_8$ & $-f_0 \ \sin\theta_0$ \\
$f_8 \ \sin\theta_8$ & $f_0 \ \cos\theta_0$ \end{tabular}\right) 
\end{eqnarray}
Phenomenological analysis of the various decays of $\eta$ and $\eta'$ to
determine $f^a_P$ has been carried out by a number of authors~\cite{hl,feld,re}.
In a recent analysis~\cite{re} Escribano and Frere find, with 
\begin{equation}
f_8 = 1.28 f_\pi \ (f_\pi = 130.7 \mbox{MeV}),
\label{para_1}
\end{equation}
the other three parameters to be
\begin{equation}
\theta_8 = (-22.2 \pm 1.8)^\circ, \hspace{0.5cm}
\theta_0 = (-8.7 \pm 2.1)^\circ, \hspace{0.5cm}
f_0=(1.18 \pm 0.04) \ f_\pi.
\label{para_2}
\end{equation}
The divergence of the axial currents are given by
\begin{eqnarray}
\partial^\mu J^8_{\mu5} &=& \frac{i \ 2}{\sqrt{6}} \left(
m_u \ \bar{u}\gamma_5 u + m_d \ \bar{d}\gamma_5 d -2 m_s \ \bar{s}\gamma_5 s
\right)\\
\partial^\mu J^0_{\mu5} &=& \frac{i \ 2}{\sqrt{3}} \left(
m_u \ \bar{u}\gamma_5 u + m_d \ \bar{d}\gamma_5 d + m_s \ \bar{s}\gamma_5 s
\right) + \frac{1}{\sqrt{3}} \ \frac{3\alpha_s}{4\pi} \ G^a_{\mu\nu}
\tilde{G}^{a\mu\nu}
\end{eqnarray}
Since $m_u,m_d << m_s$, one can neglect them~\cite{ra} to obtain
\begin{eqnarray}
\langle0|\frac{3\alpha_s}{4\pi}G^a_{\mu\nu}\tilde{G}^{a\mu\nu}|\eta\rangle &=&
\sqrt{\frac{3}{2}} \ m_\eta^2 \ \left(f_8 \ \cos\theta_8 - \sqrt{2}f_0 \
\sin\theta_0\right)\label{g2eta}\\
\langle0|\frac{3\alpha_s}{4\pi}G^a_{\mu\nu}\tilde{G}^{a\mu\nu}|\eta'\rangle &=&
\sqrt{\frac{3}{2}} \ m_{\eta'}^2 \ \left(f_8 \ \sin\theta_8 + \sqrt{2}f_0 \
\cos\theta_0\right). 
\label{g2etap}
\end{eqnarray}
Using Eqs.(\ref{vevQ}), (\ref{g2eta}) and (\ref{g2etap}) we get the
representation of $\chi(q^2)$ in terms of physical states as
\begin{eqnarray}
\chi(q^2) &=& - \frac{m_\pi^4}{8(q^2-m_\pi^2)} \ f_\pi^2 \ \left(\frac{m_d-m_u}{
m_d+m_u}\right)^2 - \frac{m_\eta^4}{24(q^2-m_\eta^2)} \ \left(f_8 \ \cos\theta_8
-\sqrt{2}f_0 \ \sin\theta_0\right)^2\nonumber\\
&&- \frac{m_{\eta'}^4}{24(q^2-m_{\eta'}^2)} \ \left(f_8 \ \sin\theta_8 + \sqrt{2}f_0 \
\cos\theta_0\right)^2 + \mbox{higher mass states.}
\label{chi-phy}
\end{eqnarray}
On the other hand, $\chi(q^2)$ has an operator product expansion~\cite{van,alk,
narison,avs}
\begin{eqnarray}
\chi(q^2)_{OPE} &=& -\left(\frac{\alpha_s}{8\pi}\right)^2 \ \frac{2}{\pi^2} \ q^4 \
\ln\left(\frac{-q^2}{\mu^2}\right) \ \left[ 1 + \frac{\alpha_s}{\pi} \ \left(
\frac{83}{4} - \frac{9}{4}\ln\left(\frac{-q^2}{\mu^2}\right)\right)\right]
\nonumber\\
&& - \frac{1}{16} \ \frac{\alpha_s}{\pi} \ \langle0|\frac{\alpha_s}{\pi} G^2 
|0\rangle \ \left(1-\frac{9}{4} \frac{\alpha_s}{\pi} \ \ln\left(\frac{-q^2}{ 
\mu^2} \right) \right) + \frac{1}{8q^2} \frac{\alpha_s}{\pi} \ \langle0|
\frac{\alpha_s}{\pi} g_s G^3 |0\rangle \nonumber\\
&&-\frac{15}{128} \frac{\pi\alpha_s}{q^4} \ \langle0|\frac{\alpha_s}{\pi} G^2
|0\rangle ^2 + 16 \left(\frac{\alpha_s}{4\pi}\right)^3 \sum_{i=u,d,s} m_i \
\langle\bar{q}_i q_i\rangle \ \left[ \ln\left(\frac{-q^2}{\mu^2}\right) +
\frac{1}{2} \right] \nonumber \\
&&-\left[ \frac{q^4}{2} \intop d\rho \ n(\rho) \ \rho^4 \ K_2^2(Q\rho) +
\mbox{screening correction to the direct instantons} \right].
\label{chi-ope}
\end{eqnarray}
In Eq.(\ref{chi-ope}), the first term arises from the perturbative gluon loop
with radiative corrections~\cite{alk}, the second, third and fourth terms are
from the vacuum expectation values of $G^2$, $G^3$ and $G^4$. The $\langle0|
G^4|0\rangle$ term has been expressed as $\langle0|G^2|0\rangle^2$ using
factorization~\cite{van}. The fifth term proportional to the quark mass has been
computed by us and is indeed quite small compared to other terms numerically.
Finally, the last two terms represent the contribution to $\chi(q^2)$ from the
direct instantons~\cite{van}. $n(\rho)$ is the density of instanton of size
$\rho$, $K_2$ is the Mc~Donald function and $Q^2=-q^2$. In a recent
work~\cite{hf}, Forkel has emphasized the importance of screening correction
which almost cancels the direct instanton contribution (cf. especially Fig.8 and
Secs.~V and VI of Ref.~\cite{hf} ). For this reason we shall disregard the
direct instanton term and screening correction for the present and return 
to it later.

From Eq.(\ref{chip0}), we now obtain
\begin{eqnarray}
\chi'(0) &=& \frac{f_\pi^2}{8} \left(\frac{m_d-m_u}{m_d+m_u}\right)^2 \left(
1+ \frac{m_\pi^2}{M^2}\right) \ e^{\frac{-m_\pi^2}{M^2}} + \frac{1}{24} 
\left( f_8 \cos\theta_8 - \sqrt{2} f_0 \sin\theta_0 \right)^2 \ \left( 1+
\frac{m_\eta^2}{M^2}\right) \ e^{\frac{-m_\eta^2}{M^2}} \nonumber \\
&&+\frac{1}{24}\left( f_8\sin\theta_8 + \sqrt{2} f_0 \cos\theta_0\right)^2 \ 
\left( 1+\frac{m^2_{\eta'}}{M^2} \right) \ e^{\frac{-m_{\eta'}^2}{M^2}}
\nonumber\\
&&-\left(\frac{\alpha_s}{4\pi}\right)^2 \frac{1}{\pi^2} \ M^2 \ E_0(W^2/M^2)
\left[1+ \frac{\alpha_s}{\pi} \frac{74}{4} + \frac{\alpha_s}{\pi} \frac{9}{2}
\left( \gamma - \ln \frac{M^2}{\mu^2}\right)\right] \nonumber\\
&&-16 \left(\frac{\alpha_s}{4\pi}\right)^3 \ \frac{1}{M^2} \ \sum_{i=u,d,s}
m_i \langle\bar{q}_i q_i\rangle - \frac{9}{64} \frac{1}{M^2} \left(
\frac{\alpha_s}{\pi}\right)^2 \left\langle \frac{\alpha_s}{\pi} G^2\right\rangle
\nonumber \\
&&+\frac{1}{16} \frac{1}{M^4} \frac{\alpha_s}{\pi} \left\langle g_s
\frac{\alpha_s}{\pi} G^3 \right\rangle - \frac{5}{128} \frac{\pi^2}{M^6}
\frac{\alpha_s}{\pi} \left\langle \frac{\alpha_s}{\pi} G^2 \right\rangle^2 .
\label{chipJP}
\end{eqnarray}
Here $E_0(x)=1-e^{-x}$ and takes into account the contribution of higher mass
states, which has been summed using duality to the perturbative term in
$\chi_{OPE}$, and $W$ is the effective continuum threshold. We take $W^2 = 2.3$
GeV$^2$, and Fig.1 plot the r.h.s. of Eq.(\ref{chipJP}) as a function of $M^2$.
We take $\alpha_s=0.5$ for $\mu=1$ GeV and 
\begin{eqnarray}
&& \langle0|g_s^2 G^2|0\rangle = 0.5 \ \mbox{GeV}^2, \  
\langle0|\bar{s}s|0\rangle = 0.8 \langle0|\bar{u}u|0\rangle  \ \mbox{with} \ 
\langle0|\bar{u}u|0\rangle = -(240 \ \mbox{MeV})^3, \nonumber\\
&& m_s = 150 \ \mbox{MeV and } \ m_u/m_d \approx 0.5.
\end{eqnarray}
Writing 
\begin{equation}
\langle0|g_s^3 \ G^3|0\rangle = \frac{\epsilon}{2} \langle0|g_s^2 \
G^2|0\rangle,
\end{equation}
as in Ref.~\cite{avs}, we take $\epsilon=1$ GeV$^2$. We also have the 
PCAC relation,
\begin{equation}
-2(m_u+m_d) \ \langle0|\bar{u}u|0\rangle = f_\pi^2 \ m_\pi^2.
\end{equation}
For $f_0, \ f_8, \ \theta_8$ and $\theta_0$ we use the central values given in
Eqs.(\ref{para_1}) and (\ref{para_2}).

\begin{figure}
\centerline{\epsfig{height=10cm,file=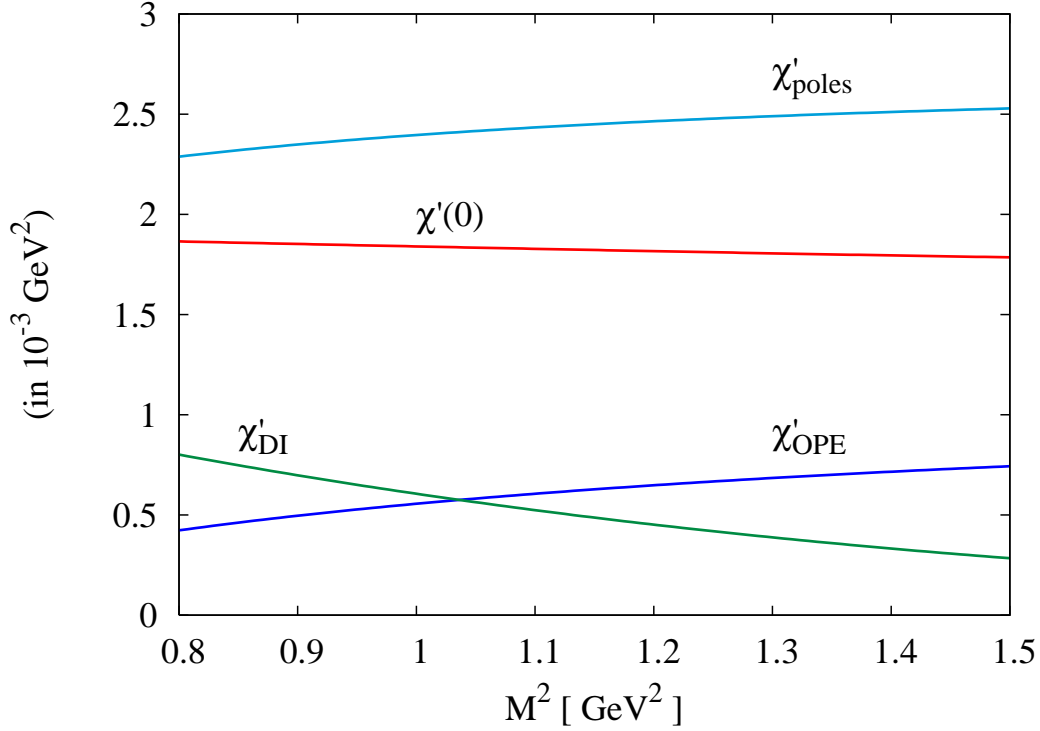}}
\caption{Various ters contributing to $\chi'(0)$, Eq.(\ref{chipJP}). The value
of $\chi'(0)$ is the one obtained without the direct instantons. The latter, see
Eq.(29), is given by $\chi'_{DI}$, which is larger than $\chi'_{OPE}$ and also
has the wrong $M^2$ behaviour suggesting that screening corrections are
important.}
\end{figure}
Let us now examine how the various terms in the r.h.s. of Eq.(\ref{chipJP}) add
up to remain a constant. The pion term is small and has little variation because
of its low mass, $\eta$ and $\eta'$ are significantly larger and $\eta$ is even
larger that $\eta'$. In Fig.1 the upper line gives the combined contribution of
$\pi, \ \eta$ and $\eta'$ which we denote as $\chi'_{poles}$ and it is seen
that it has a gentle increase with $M^2$. The OPE term given by the last three
lines in Eq.(\ref{chipJP}), which we denote by $\chi'_{OPE}$, so that 
$$\chi'(0) = \chi'_{poles} - \chi'_{OPE}$$
is also plotted in Fig.1. It is seen that $\chi'_{OPE}$ is roughly about 25\%
of $\chi'_{poles}$ also increases with $M^2$, with the result that
$\chi'(0)$ is nearly a constant w.r.t $M^2$.

We expect this trend of compensating variation in $\chi'_{poles}$ and
$\chi'_{OPE}$ to be maintained when variation in $\chi'_{poles}$ due to
uncertainties in $\theta_8, \ \theta_0, \ f_8, \ f_0$ [see Eqs.(\ref{para_1})]
and (\ref{para_2}) and the variations in $\chi'_{OPE}$ due to uncertainties in
the estimates of the vacuum condensates are taken into account.
We can then obtain, from Fig.1, the value
\begin{equation}
\chi'(0) \approx 1.82 \times 10^{-3} \ \mbox{GeV}^2.
\label{Nchip0}
\end{equation}
We note that the above determination, Eq.(\ref{Nchip0}), is in agreement with 
an entirely different calculation by two of us~\cite{jp} from the study of the
correlator of isoscalar axial vector currents
\begin{eqnarray}
\Pi_{\mu\nu}^{I=0} &=& \frac{i}{2} \intop d^4x \ e^{i q.x} \langle0| \left\{
\bar{u}\gamma_\mu\gamma_5u(x) + \bar{d}\gamma_\mu\gamma_5d(x), 
\bar{u}\gamma_\mu\gamma_5u(0) + \bar{d}\gamma_\mu\gamma_5d(0) \right\}|0\rangle 
\nonumber\\
\Pi_{\mu\nu}^{I=0} &=& -\Pi_{1}^{I=0}(q^2) \ g_{\mu\nu} + \Pi_{2}^{I=0}(q^2)
q_\mu q_\nu.
\end{eqnarray}
$\Pi_1^{I=0}(q^2=0)$ can be computed from the spectrum of axial vector mesons.
In Ref.~\cite{jp} a value
\begin{equation}
\Pi_1^{I=0}(q^2=0) = -0.0152 \ \mbox{GeV}^2
\label{Pi0}
\end{equation}
was obtained. It is not difficult to see that when $m_u=m_d=0$
\begin{equation}
\chi'(0) = - \ \frac{1}{8} \ \Pi_1^{I=0}(q^2=0)
\end{equation}
which is consistent with Eqs.(\ref{Nchip0}).
Let us now return to Eq.(\ref{chi-ope}) and consider the effect of
incorporating the direct instanton term in Eq.(\ref{chipJP}) in the 
spike approximation~\cite{avs}:
\begin{equation}
n(\rho) = n_0 \ \delta(\rho - \rho_c)
\end{equation}
with $n_0=0.75 \times 10^{-3}$ GeV$^4$ and $\rho_c = 1.5$ GeV$^{-1}$. The
contribution of the direct instanton to $\hat{B}[\chi'(q^2)/q^2]$ can be 
found using the asymptotic expansion for $K_2(z)$ and $K_2'(z)$ and we find
it to be 
\begin{equation}
\chi'_{DI} = 
\frac{n_0}{4} \ \sqrt{\pi} \ \rho_c^4 \ M^2 \left[ M\rho_c + \frac{9}{4}
\frac{1}{M \rho_c} + \frac{45}{32} \frac{1}{M^3\rho_c^3} \right] \ 
e^{-M^2\rho_c^2}.
\label{DI}
\end{equation}
We have plottend this term separately in Fig.1. We note that unlike
$\chi'_{poles}$ and $\chi'_{OPE}$, which increase with $M^2$ and therefore
compensate each other, the contribution of $\chi'_{DI}$, Eq.(\ref{DI}),
decreases rapidly with $M^2$. It is not difficult to see that 
$\chi'(0)$ will no longer remain constant. This strongly suggests that 
screening corrections to $\chi'_{DI}$ are important just as they are for 
$[\chi(q^2)/q^2]$ as found by Forkel~\cite{hf}.

We now turn to an estimate of $\eta'$ mass in the chiral limit: $m_u = m_d = m_s
= 0$. In this limit $SU(3)$ flavor symmetry is exact and, we have $m_\pi= m_\eta
= 0$ while $\eta'$ is a singlet. Let us denote by $\eta_\chi = \eta'(m_s=0)$ and
$m_\chi = m_{\eta'}(m_s=0)$, the singlet particle and its mass in the chiral
limit. Returning to Eq.(\ref{chi-ope}), we first note that the explicitly quark
mass dependent term in $\chi_{OPE}$
\[ -16\left(\frac{\alpha_s}{4\pi}\right)^3 \ \sum_{i=u,d,s} m_i \langle 
\bar{q}_i q_i \rangle \approx 1.9 \times 10^{-6} \ \mbox{GeV}^4 \]
is numerically much smaller than, for expample
\[ \frac{9}{64}\left(\frac{\alpha_s}{\pi}\right)^2 \left\langle \frac{\alpha_s}
{\pi} G^2\right\rangle \approx 4.5 \times 10^{-5} \ \mbox{GeV}^4 \]
which itself is much smaller than the perturbative term. In the chiral limit
$\langle0|Q|\pi\rangle = \langle0|Q|\eta\rangle = 0$. If we assume that the
quark mass dependence of $\chi'(0)$ is negligible then $\chi'(0)$ in
Eq.(\ref{chipJP}) can also be expressed in term of $f_{\eta_\chi}$ and $m_\chi$
as:
\begin{equation}
\chi'(0) = \frac{1}{12} f_{\eta_\chi}^2 \ \left( 1+ \frac{m_\chi^2}{M^2}\right)
e^{\frac{-m_\chi^2}{M^2}} - \hat{B}\left[\frac{\chi'_{OPE}(q^2)}{q^2}\right].
\label{chipJP1}
\end{equation}
From Eqs.(\ref{chipJP}) and (\ref{chipJP1}) we may then write 
\begin{eqnarray}
\frac{1}{12}f_{\eta_\chi}^2 \ \left( 1+ \frac{m_\chi^2}{M^2}\right)
e^{\frac{-m_\chi^2}{M^2}} & \approx &
\frac{1}{24}f_{\pi}^2 \ \left( 1+ \frac{m_\pi^2}{M^2}\right)
e^{\frac{-m_\pi^2}{M^2}}\nonumber\\
&+& \frac{1}{24} \left(f_8\cos\theta_8 - \sqrt{2} f_0 \sin\theta_0 \right)^2
\left( 1 + \frac{m_\eta^2}{M^2} \right)e^{\frac{-m_\eta^2}{M^2}}\nonumber\\
&+& \frac{1}{24} \left(f_8\sin\theta_8 + \sqrt{2} f_0 \cos\theta_0 \right)^2
\left( 1 + \frac{m_{\eta'}^2}{M^2} \right)e^{\frac{-m_{\eta'}^2}{M^2}}.
\label{chiLR}
\end{eqnarray}
We can find $f_{\eta_\chi}$ and $m_\chi$ from Eq.(\ref{chiLR}) using the least
``chi-squared" criterion in the range 0.8 GeV$^2 < M^2 <$ 1.5 GeV$^2$. We find
$m_\chi \approx 723$ MeV and corresponding $f_{\eta_\chi} = 178$ MeV.
In Fig.2 we have plotted the l.h.s. and r.h.s. of Eq.(\ref{chiLR}) as a function
of $M^2$ for best-fit values of $m_\chi$ and $f_{\eta_\chi}$.
The decay
constant $f_{\eta_\chi}$ is of the same order as physical decay constants $f_8$
and $f_0$.

\begin{figure}
\centerline{\epsfig{height=10cm,file=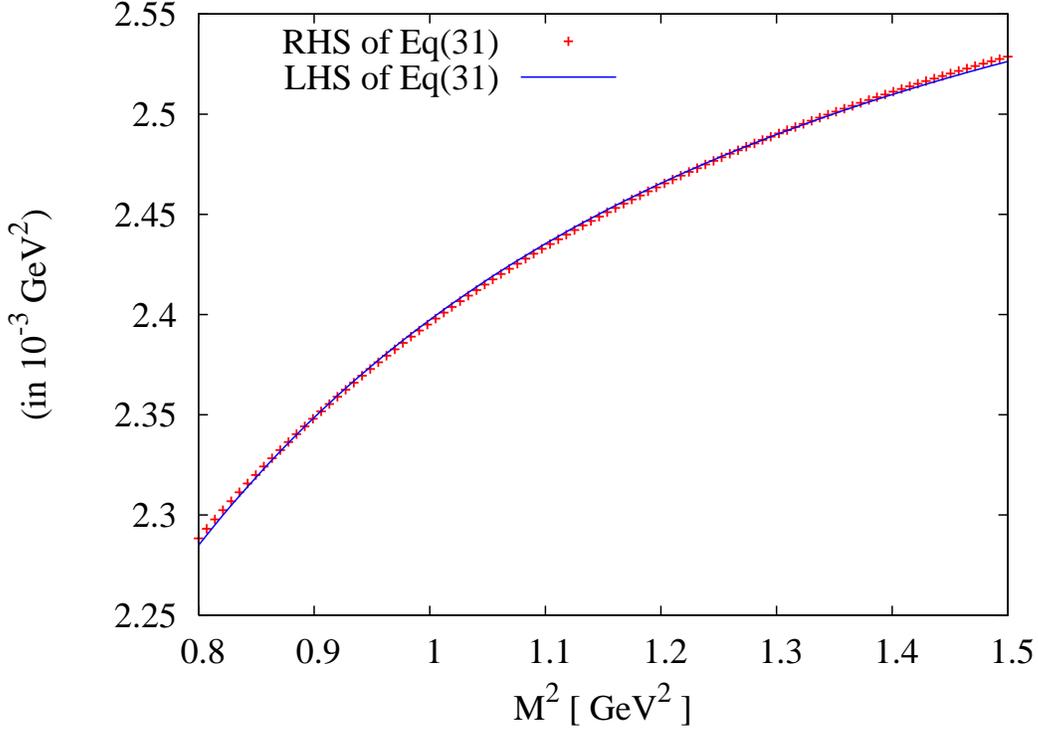}}
\caption{Estimate of $\eta'$ mass and coupling in the chiral limit, see Eq.(31). 
The continuous curve corresponds to $m_\chi = 723$ MeV.}
\end{figure}
We now compare our result for $\chi'(0)$ with some earlier results. In
Ref.~\cite{narison}, Narison {\it et al} obtained a value for $\chi'(0)
\approx 0.7 \times 10^{-3}$ GeV$^2$, substantially different from the value
derived here. Since the expression for $\chi_{OPE}$ used by us is identical to
theirs, albeit the estimate used for the gluon condensates are slighly 
different, we need to explain the difference in the end result for $\chi'(0)$.
The most important difference is
in the expression of $\chi(q^2)$ in terms of physical intermediate states. We
have seen that both $\eta$ and $\eta'$ contribute, and in fact $\eta$ makes a
larger contribution than $\eta'$. In Ref.~\cite{narison} only $\eta'(958)$
state is taken into account. We have also seen that if we were to take the
chiral limit then $\eta$ and $\eta'$ contribution to $\chi(q^2)$ is
representable by $\eta_\chi$ with mass $m_\chi = 723$ MeV, which is
substantially different form the physical $\eta'$ mass. This also explains why
Narison {\it et al.} find stability in the sum rule only for rather larger 
$W^2$(=6 GeV$^2$) instead of our $W^2$=2.3 GeV$^2$. We must also add that while
our Eq.(\ref{chip0}) involves only $[\chi'(q^2)/q^2]$, Narison {\it et al.} 
use the linear combination of two sum rules (cf. Eq.(6.22) of 
Ref.~\cite{narison}).

In Ref.~\cite{ayk}, Ioffe and Khodzhamiryan's claim that the OPE for $\chi(q^2)$
does not converge is based on the following. They computed the correlators
\[ i \ q_\mu q_\nu \intop d^4x \ e^{iq.x} \langle0| T\left\{J^0_{\mu5}(x),
J^q_{\nu5}(0) \right\} |0\rangle \]
where, $J^q_{\mu5} = \bar{q}\gamma_\mu\gamma_5 q \ (q = u,d,s)$ with $m_u=m_d =0$ but $m_s
\neq 0$ and $j^0_{\mu5}$ is the flavor singlet axial current. Introducing the
definition
\[ \langle0|J^q_{\mu5}(x) | \eta'(p) \rangle = i p_\mu g^q_{\eta'} \]
they estimated 
\begin{equation}
 g^s_{\eta'}/g^u_{\eta'} \approx 2.5 . 
 \label{ratio}
\end{equation}
If $SU(3)$ symmetry were
to be exact, this ratio would be unity. Insisting that the ratio in
Eq.(\ref{ratio}) should be close to unity even when $m_s \neq 0$, they
concluded that their result signals a breakdown of OPE~\cite{ayk}. As
discussed earlier, $\langle0|J^8_{\mu5}|\eta'\rangle \neq 0$. In fact using the
values given in Eqs.(\ref{para_1}) and (\ref{para_2}), it is easy to obtain
\begin{equation}
\frac{g^s_{\eta'}}{g^u_{\eta'}} = \frac{\sqrt{2} (f_0 \cos_0 - \sqrt{2} f_8
\sin\theta_8)}{f_8 \sin_8 + \sqrt{2} f_0 \cos\theta_0)} \approx 2.24
\end{equation}
which is close enough to the estimate of Ref.~\cite{ayk}. As in the case of
Narison {\it et al}~\cite{narison}, Ioffe and Samsonov~\cite{avs} and
Forkel~\cite{hf} also do not take into account the $\pi, \ \eta$ matrix element
of the anomaly in their sum rules involving $\chi(q^2)$.

In conclusion we find a value of $\chi'(0) \approx 1.82 \times 10^{-3}$ GeV$^2$
without incorporating direct instantons. Screening corrections to the latter
appears to be significant. We also obtain an estimate $m_\chi = 723$ MeV and
$f_{\eta_\chi}=178$ MeV.

{\bf ACKNOWLEDGEMENT:} JPS and AU thank CHEP, IISc,
Bangalore, for their hospitality at the Center where part of this work was done.

\end{document}